\documentclass[conference]{IEEEtran}
\ifCLASSINFOpdf
\else
\fi
\usepackage{times} 
\usepackage{graphicx}                              
\usepackage{epsfig}
\usepackage{grffile}
\usepackage{algorithmic}
\usepackage{amssymb}
\usepackage{amsmath}
\usepackage[ruled]{algorithm}
\usepackage{slashbox}
\usepackage{stmaryrd}

\newtheorem{lemma}{Lemma}[section]  

\newcommand{\reffig}[1]{Figure~\ref{#1}}
\newcommand{\reffigwo}[1]{\ref{#1}}

\newcommand{\reflem}[1]{Lemma~\ref{#1}}

\newcommand{\refeq} [1]{equation~(\ref{#1})}
\newcommand{\refeqwo}[1]{(\ref{#1})}

\newcommand{\bs}[1]{{\boldsymbol{#1}}}
\newcommand{\sym}{{{\psi}}}

\begin{document}
%
\title{Trellis-Based Check Node Processing for Low-Complexity Nonbinary LP Decoding}
%
\author{Mayur Punekar and Mark F. Flanagan\\
Claude Shannon Institute, \\
University College Dublin, Belfield, Dublin 4, Ireland. \\
{\{mayur.punekar, mark.flanagan\}@ieee.org}
}
%
\maketitle
%
\begin{abstract}
Linear Programming (LP) decoding is emerging as an attractive alternative to decode Low-Density Parity-Check (LDPC) codes. However, the earliest LP decoders proposed for binary and nonbinary LDPC codes are not suitable for use at moderate and large code lengths.
To overcome this problem, Vontobel \textit{et al.} developed an iterative Low-Complexity LP (LCLP) decoding algorithm for binary LDPC codes. 
The variable and check node calculations of binary LCLP decoding algorithm are related to those of binary Belief Propagation (BP).
The present authors generalized this work to derive an iterative LCLP decoding algorithm for nonbinary linear codes. Contrary to binary LCLP, the variable and check node calculations of this algorithm are in general different from that of nonbinary BP. The overall complexity of nonbinary LCLP decoding is linear in block length; however the complexity of its check node calculations is exponential in the check node degree. In this paper, we propose a modified BCJR algorithm for efficient check node processing in the nonbinary LCLP decoding algorithm. The proposed algorithm has complexity linear in the check node degree.
We also introduce an alternative state metric to improve the run time of the proposed algorithm. Simulation results are presented for $(504, 252)$ and $(1008, 504)$ nonbinary LDPC codes over $\mathbb{Z}_4$.
\end{abstract}
%
\vspace{-5pt}
\section{Introduction}
Binary and nonbinary LDPC codes \cite{DaMa_98} have attracted much attention in the research community in the past decade. LDPC codes are generally decoded by the iterative BP algorithm which performs remarkably well at moderate SNR levels. 
Due to their capacity achieving performance,
LDPC codes are used in many current communications systems. They are also a promising candidate for future high data rate communication systems as well as for memory applications. 
However, BP suffers from a so called \textit{error floor} problem at high SNR. Also, the heuristic nature of BP makes it difficult to analyze, and simulations are too time consuming for the prediction of the error floor.

In recent years, the new approach of LP decoding is emerging as an attractive alternative to the BP decoding. LP decoding for binary LDPC codes was proposed by Feldman \textit{et al.} \cite{FeWa_05}. In LP decoding, the maximum likelihood decoding problem is modeled as an LP problem. 
In contrast to BP decoding, LP decoding relies on a well studied branch of mathematics which provides a basis for better understanding of the decoding algorithms. The work of \cite{FlSk_09} extended the LP decoding framework of Feldman \textit{et al.} to nonbinary linear codes. Binary and nonbinary LP decoding algorithms rely on standard LP solvers based on simplex or interior point methods. However, the time complexity of these solvers is known to be exponential in number of variables, which limits the use of LP decoding to codes of small block length.
To decode longer codes, a specialized low complexity LP decoding algorithm is necessary. Such a low-complexity algorithm for binary LDPC codes was proposed by Vontobel \textit{et al.} in \cite{VoKo_06}. The present authors, in \cite{PuFl_10}, extended the binary LCLP decoding algorithm \cite{VoKo_06} to nonbinary codes. The complexity of the proposed nonbinary LCLP decoding algorithm is linear in the block length. As opposed to binary LCLP decoding, nonbinary LCLP decoding is not directly related to nonbinary BP. Due to this, the complexity of the check node calculations of nonbinary LCLP decoding is exponential in the maximum check node degree. In this paper, we propose a modified BCJR algorithm for the check node processing of nonbinary LCLP decoding. The proposed algorithm has complexity linear in the check node degree and allows for efficient implementation of nonbinary LCLP decoding. We also propose an alternative state metric which can be used for faster check node processing.

This paper is organized as follows. Notation and background information is given in Section II. Section III reviews the nonbinary LCLP decoding algorithm from \cite{PuFl_10}. Section IV contains the modified BCJR algorithm for check node processing and also explains the alternative state metric. Section V presents the simulation results, and Section VI concludes the paper.
\section{Notation and Background}
Let $\Re$ be a finite ring with $q$ elements with $0$ as its additive identity. We define $\Re^{-} = \Re \setminus \{0\}$. Let $\mathcal{C}$ be a linear code of length $n$ over the ring $\Re$, defined by
$\quad\quad\quad\quad\quad\quad\quad\quad\quad\quad\quad\mathcal{C} = \{ \bs{c} \in \Re^{n} : \bs{c} \mathcal{H}^{T} = \bs{0} \}$,
where $\mathcal{H}$ is a $m \times n$ parity-check matrix with entries from $\Re$. $R(\mathcal{C}) = \log_q (|\mathcal{C}|) / n$ is the rate of code $\mathcal{C}$. Hence, the code $\mathcal{C}$ is an $[n, \log_q(|\mathcal{C}|)]$ linear code over $\Re$. The row indices and column indices of $\mathcal{H}$ are denoted by the sets $\mathcal{J} = \{1,\ldots,m\}$ and $\mathcal{I} = \{1,\ldots,n\}$ respectively. The $j$-th row of $\mathcal{H}$ is denoted by $\mathcal{H}_j$ and the $i$-th column of $\mathcal{H}$ is denoted by $\mathcal{H}^i$. supp$(\bs{c})$ denotes the support of the vector $\bs{c}$. For each $j \in \mathcal{J}$, let $\mathcal{I}_j = \mbox{supp}(\mathcal{H}_j)$ and for each $i \in \mathcal{I}$, let $\mathcal{J}_i = \mbox{supp}(\mathcal{H}^i)$. Also let $d_j = |\mathcal{I}_j|$ and $d = \max_{j \in \mathcal{J}}\{d_j\}$. We define the set $\mathcal{E} = \{(i,j) \in \mathcal{I} \times \mathcal{J} \; : \; j \in \mathcal{J}, i \in \mathcal{I}_j\} = \{(i,j) \in \mathcal{I} \times \mathcal{J} \; : \; i \in \mathcal{I}, j \in \mathcal{J}_i\}$. Moreover for each $j \in \mathcal{J}$ we define the local Single Parity Check (SPC) code 
\begin{equation*}
\mathcal{C}_j = \left \{(b_i)_{i \in \mathcal{I}_j} : \sum_{i \in \mathcal{I}_j} b_i \cdot \mathcal{H}_{j,i} = 0 \right\} 
\end{equation*}
For each $i \in \mathcal{I}$, $\mathcal{A}_i \subseteq \Re^{|\{0\} \cup \mathcal{J}_i|}$ denotes the repetition code of the appropriate length and indexing. We also use variables $\bs{u}_{i,j} = (u_{i,j}^{(\alpha)})_{\alpha \in \Re^{-}}$ and $\bs{v}_{j,i} = (v_{j,i}^{(\alpha)})_{\alpha \in \Re^{-}}$ for all $i \in \mathcal{I}$, $j \in \mathcal{J}_i \cup \{ 0 \}$; also for $i \in \mathcal{I}$, $\bs{u}_i = (\bs{u}_{i,j})_{j \in \mathcal{J}_i \cup \{ 0 \}}$ and similarly for $j \in \mathcal{J}$, $\bs{v}_j = (\bs{v}_{j,i})_{i \in \mathcal{I}_j}$.

We use the following mapping given in \cite{FlSk_09},
\begin{equation*}
\xi : \Re \rightarrow \{0,1\} ^{q-1} \subset \mathbb{R}^{q-1}
\end{equation*}
by
\begin{equation*}
\xi(\alpha) = \bs{x} = (x^{(\rho)})_{\rho \in \Re^{-}}
\end{equation*}
such that, for each $\rho \in \Re^{-}$
\begin{eqnarray*}
x^{(\rho)} = \left\{ 
\begin{array}{l l}
  1, & \; \text{if} \; \rho = \alpha \\
  0, & \; \text{otherwise}\\
\end{array} \right.
\end{eqnarray*}
We extend this mapping to define
\[
\Xi : \underset{t \in \mathbb{Z}^{+}}{\cup} \Re^{t} \rightarrow \underset{t \in \mathbb{Z}^{+}}{\cup} \{0,1\}^{(q-1)t} \subset \underset{t \in \mathbb{Z}^{+}}{\cup} \mathbb{R}^{(q-1)t} \; ,
\]
where, 
\[
\Xi (\bs{c}) = (\xi(c_1),\dots,\xi(c_t))\nonumber, \quad \forall \bs{c} \in \Re^{t}, t \in \mathbb{Z}^{+} \; .
\]

For $\kappa \in \mathbb{R}, \kappa > 0$, we define the function $\sym (x) = e^{\kappa x}$ and its inverse $\sym^{-1} (x) = \frac{1}{\kappa} \log (x).$
We also use the soft-minimum operator introduced in \cite{VoKo_06}. For any $\kappa \in \mathbb{R}$, $\kappa > 0$, the soft-minimum operator is defined as 
\begin{align*}
\min_{l}{}^{(\kappa)} \{ z_l \} \triangleq -\frac{1}{\kappa} \log \left( \sum_{l} e^{-\kappa z_l}\right) = -\sym^{-1}\left(\sum_{l}\sym\Big({-z_l}\Big)\right)
\end{align*}
where $\min_{l}{}^{(\kappa)} \{ z_l \} \le \min_{l} \{z_{l}\}$ with equality attained in the limit as $\kappa \to \infty$.  

We assume transmission over a $q$-ary input memoryless channel and also assume a corrupted codeword $\bs{y} =(y_1, y_2, \cdots, y_n)\in \Sigma^{n}$ has been received. Here, the channel output symbols are denoted by $\Sigma$. Based on this, we define a vector $\bs{\lambda} = (\lambda^{(\alpha)})_{\alpha \in \Re^{-}} $ 
where, for each $y \in \Sigma$, $\alpha \in \Re^{-}$,
\[
\lambda^{(\alpha)} = \log \left(\frac{p(y|0)}{p(y|\alpha)}\right) \; .
\]
Here $p(y|c)$ denotes the channel output probability (density) conditioned on the channel input. 
%
%
%
%
%
%
%
%
\section{Low Complexity LP Decoding of Nonbinary Linear Codes}
To develop a low complexity LP solver for nonbinary linear codes, the present authors in \cite{PuFl_10} proposed a primal LP formulation which is equivalent to the original LP formulation proposed in \cite{FlSk_09}. This primal LP formulation has an advantage that, it has one-to-one corresponding with the Forney-style factor graph of the code and can be used to derive a suitable dual LP (see section IV in \cite{PuFl_10}). The dual LP is then ``softened'' by using the ``soft-min'' operator which is used to derive the update equations given in \textit{Lemma 6.1} in \cite{PuFl_10}. The softened dual LP is given below.
\begin{eqnarray}
\textbf{SDNBLPD:} && \nonumber \\
\text{max.}	&& \sum_{i \in \mathcal{I}} \hat{\phi}_{i} + \sum_{j \in \mathcal{J}} \hat{\theta}_{j} \nonumber \\
\mbox{Subject to } && \nonumber \\
\hat{\phi}_{i} &\leq& \min_{\bs{a} \in \mathcal{A}_i}{}^{(\kappa)} \left\langle-\hat{\bs{u}}_i, \Xi(\bs{a})\right\rangle \quad (i \in \mathcal{I}), \nonumber \\
\hat{\theta}_{j} &\leq& \min_{\bs{b} \in \mathcal{C}_j}{}^{(\kappa)} \left\langle-\bs{\hat{v}}_j, \Xi(\bs{b})\right\rangle \quad  (j \in \mathcal{J}), \nonumber \\
\hat{\bs{u}}_{i,j} &=& - \hat{\bs{v}}_{j,i}\quad \quad \quad \quad \quad \quad((i,j) \in \mathcal{E}), \nonumber \\
\hat{\bs{u}}_{i,0} &=& - \hat{\bs{f}}_i \quad \quad \quad \quad \quad \quad \quad (i \in \mathcal{I}), \nonumber \\
\hat{\bs{f}}_i &=& \bs{\lambda}_i \quad \quad \quad \quad \quad \quad \quad (i \in \mathcal{I}). \nonumber 
\end{eqnarray}
The update equation can be used to update the dual variable $\hat{u}_{i,j}^{(\alpha)}$ related to an edge $(i,j) \in \mathcal{E}$ while all other edge variables are held constant. The updated value of the $\hat{u}_{i,j}^{(\alpha)}$ is given by
\begin{equation*}
\quad \bar{u}_{i,j}^{(\alpha)} = \frac{1}{2} \left( (V_{i,\bar{\alpha}} - V_{i, \alpha}) - (C_{j,\bar{\alpha}} - C_{j,\alpha}) \right)
\end{equation*}
where,
\begin{align*}
&V_{i,\bar{\alpha}} \triangleq - \min_{\underset{a_j \ne \alpha}{\bs{a} \in \mathcal{A}_i}} {}^{(\kappa)} \left\langle -\hat{\bs{u}}_i, \Xi({\bs{a}}) \right\rangle, \\
&V_{i,\alpha} \triangleq - \min_{\underset{a_j = \alpha}{\bs{a} \in \mathcal{A}_i}} {}^{(\kappa)} \left\langle -\tilde{\bs{u}}_i, \Xi(\tilde{\bs{a}}) \right\rangle, \\
&C_{j,\bar{\alpha}} \triangleq - \min_{\underset{b_i \ne \alpha}{\bs{b} \in \mathcal{C}_j}} {}^{(\kappa)} \left\langle -\hat{\bs{v}}_j, \Xi(\bs{b}) \right\rangle, \\
&C_{j,\alpha} \triangleq - \min_{\underset{b_i = \alpha}{\bs{b} \in \mathcal{C}_j}} {}^{(\kappa)} \langle -\tilde{\bs{v}}_j, \Xi(\tilde{\bs{b}}) \rangle. 
\end{align*}
Here the vector $\tilde{\bs{u}}_{i}$ is the vectors $\hat{\bs{u}}_i$ where the subvector $\hat{\bs{u}}_{i,j}$ is excluded. Similarly vector $\tilde{\bs{v}}_{j}$ is obtained by excluding the subvector $\hat{\bs{v}}_{j,i}$ from $\hat{\bs{v}}_j$. Vector $\tilde{\bs{a}}$ is same as ${\bs{a}}$ where the $j$-th position is omitted and vector $\tilde{\bs{b}}$ is obtained by excluding the $i$-th position from ${\bs{b}}$.
Now by updating all the edges $(i,j) \in \mathcal{E}$ with some schedule (e.g. circular), the low-complexity LP decoding algorithm converges to the maximum of the \textbf{SDNBLPD}. (see \textit{Lemma 6.2} in \cite{PuFl_10}). The overall complexity of this algorithm is linear in the block length. 

The terms $(V_{i,\bar{\alpha}} - V_{i, \alpha}) \text{ and } (C_{j,\bar{\alpha}} - C_{j,\alpha})$ are related to the variable node (VN) $i \in \mathcal{I}$ and check node (CN) $j \in \mathcal{J}$ respectively.
In the binary case, these terms can be efficiently calculated with the VN and CN calculations of the binary Sum-Product (SP) algorithm respectively \cite{VoKo_06}. However, for nonbinary codes, the calculation of $(V_{i,\bar{\alpha}} - V_{i, \alpha}) \text{ and } (C_{j,\bar{\alpha}} - C_{j,\alpha})$ is not related to the VN and CN calculations of the nonbinary SP algorithm \cite{PuFl_10}. Hence, the CN calculations are carried out by processing exhaustively all of the possible codewords of the SPC code $\mathcal{C}_j$. Consequently, the complexity of calculating $(C_{j,\bar{\alpha}} - C_{j,\alpha})$ (i.e of CN calculation) is in exponential in the maximum check-node degree $d$.
%
\section{Modified BCJR algorithm for Check Node calculation of the Low Complexity LP Decoding}
In \cite{PuFl_10}, the authors suggested that the equations for $C_{j,\bar{\alpha}}$ and $C_{j,\alpha}$ can be rewritten as follows:
\begin{align}
&\sym \Big(C_{j,\bar{\alpha}}\Big) = \sum_{\underset{b_i \ne \alpha}{\bs{b} \in \mathcal{C}_j}} \sym \Big(\left\langle \hat{\bs{v}}_j, \Xi(\bs{b}) \right\rangle \Big) \label{eq:c_alpha_bar}\\
& \sym \Big(C_{j,\alpha}\Big) = \sum_{\underset{b_i = \alpha}{\bs{b} \in \mathcal{C}_j}} \sym \Big(\langle \tilde{\bs{v}}_j, \Xi(\tilde{\bs{b}}) \rangle\Big) \label{eq:c_alpha}
\end{align}

\vspace{-5pt}
It may be observed from the above equations that the calculation of the $C_{j,\bar{\alpha}} \; \text{and} \; C_{j,\alpha}$ is in the form of the marginalization of a product of functions. Hence it is possible to compute $C_{j,\bar{\alpha}} \text{ and } C_{j,\alpha}$ with the help of a trellis based variant of the SP algorithm (i.e. BCJR-type algorithm). One possibility is to use the trellis of the binary nonlinear code $\mathcal{C}^{NL}_j = \{\Xi(\bs{b}) : \forall \bs{b} \in \mathcal{C}_j\}$. However, due to nonlinear nature of this binary code, the state complexity at the center of its trellis would be exponential in $d_j$. Here state merging is also not possible. Hence there is no complexity advantage when we use the trellis of the binary nonlinear code $\mathcal{C}^{NL}_j$.

However if the trellis for the nonbinary SPC code $\mathcal{C}_j$ is used, then the state complexity at each trellis step is $\mathcal{O}(q)$ and is independent of $d_j$. The branch complexity of this trellis is $\mathcal{O}(q^2)$. In the following, we prove that the marginals $C_{j,\bar{\alpha}}$ and $C_{j,\alpha}$ can be efficiently calculated with some modifications to the BCJR algorithm which uses the trellis of the nonbinary code $\mathcal{C}_j$. For this purpose we define the following for the trellis of the SPC code $\mathcal{C}_j$:
\begin{enumerate}
\item The set of all states at time $t, \; \mathcal{S}_t, t \in (0, \cdots, d_j)$ 
\item $(s, s') \in (\mathcal{S}_t, \mathcal{S}_{t+1})$ represents a branch in the trellis which is related to the symbol $b_t = s'-s$.
\item Since we have trellis for SPC code, each state $s \in \mathcal{S}_t$ represents the sum of all symbols from $b_0$ to $b_{t-1}$.
\item We define
\vspace{-10pt}
\begin{align*}
&\sigma(i, j) = \sum_{r = i}^{r=j} \;\; b_r, \quad \bs{b} \in \mathcal{C}_j.
\end{align*}
\vspace{-5pt}
\item Branch metric for each $(s, s') \in (\mathcal{S}_t, \mathcal{S}_{t+1})$ is $g(s, s') = g(b_t) = \sym \left( \langle\hat{\bs{v}}_{j,t}\;,\; \xi(b_t)\rangle\right)$.
\item State metric for forward recursion,
\begin{align}
&\mu_i(s) = \sum_{\underset{\sigma(0, {i-1}) = s}{(b_0,\cdots,b_{i-1})}} \prod_{t = 0}^{i-1} g(b_t), \quad s \in \mathcal{S}_i, i \in \mathcal{I}_j \label{eq:mu}\\
& \text{ with } \mu_{0}({0}) = 1, \quad  \mu_{0}({\alpha}) = 0, \forall \alpha \in \Re^{-}.\nonumber
\end{align} 
and state metric for backward recursion,  
\begin{align}
&\nu_{i}({s}) = \sum_{\underset{\sigma(i,d_j-1) = s}{(b_{i},\cdots,b_{d_j-1})}} \; \prod_{t=i}^{d_j-1} g(b_t), \quad s \in \mathcal{S}_{i}, i \in \mathcal{I}_j \label{eq:nu}\\
& \text{ with } \nu_{d_j}({0}) = 1, \quad  \nu_{d_j}({\alpha}) = 0, \forall \alpha \in \Re^{-}. \nonumber
\end{align}
\end{enumerate}
\begin{figure*}
\begin{minipage}[b]{0.47\linewidth}
\centering
\includegraphics[width=0.9\columnwidth]{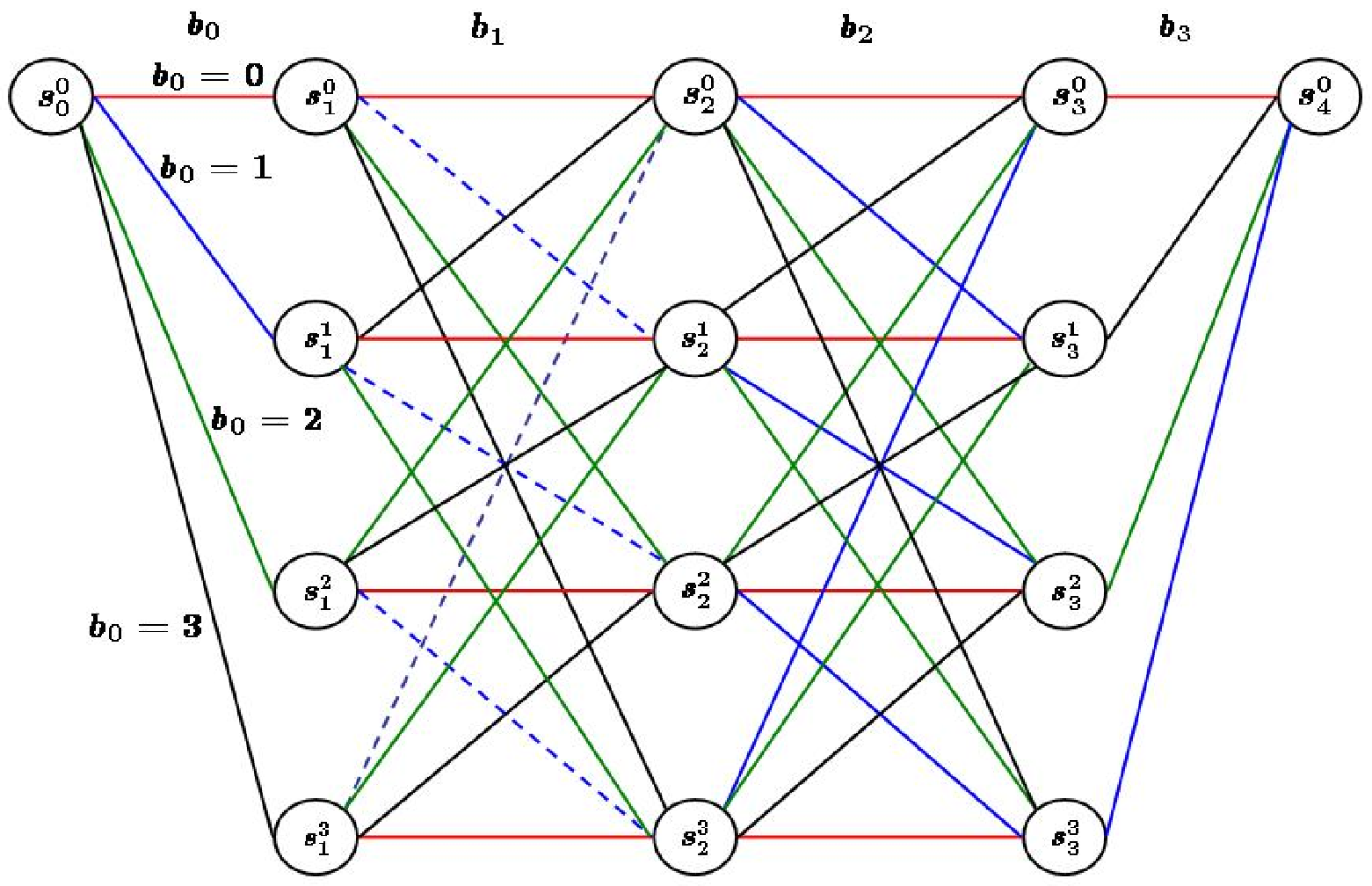}
\caption{States connected by dotted branches are used for the calculation of the $C_{j,1}$.} 
\vspace{-15pt}
\label{fig:trellis_alpha}
\end{minipage}
\hspace{1cm}
\begin{minipage}[b]{0.47\linewidth}
\centering 
\includegraphics[width=0.9\columnwidth]{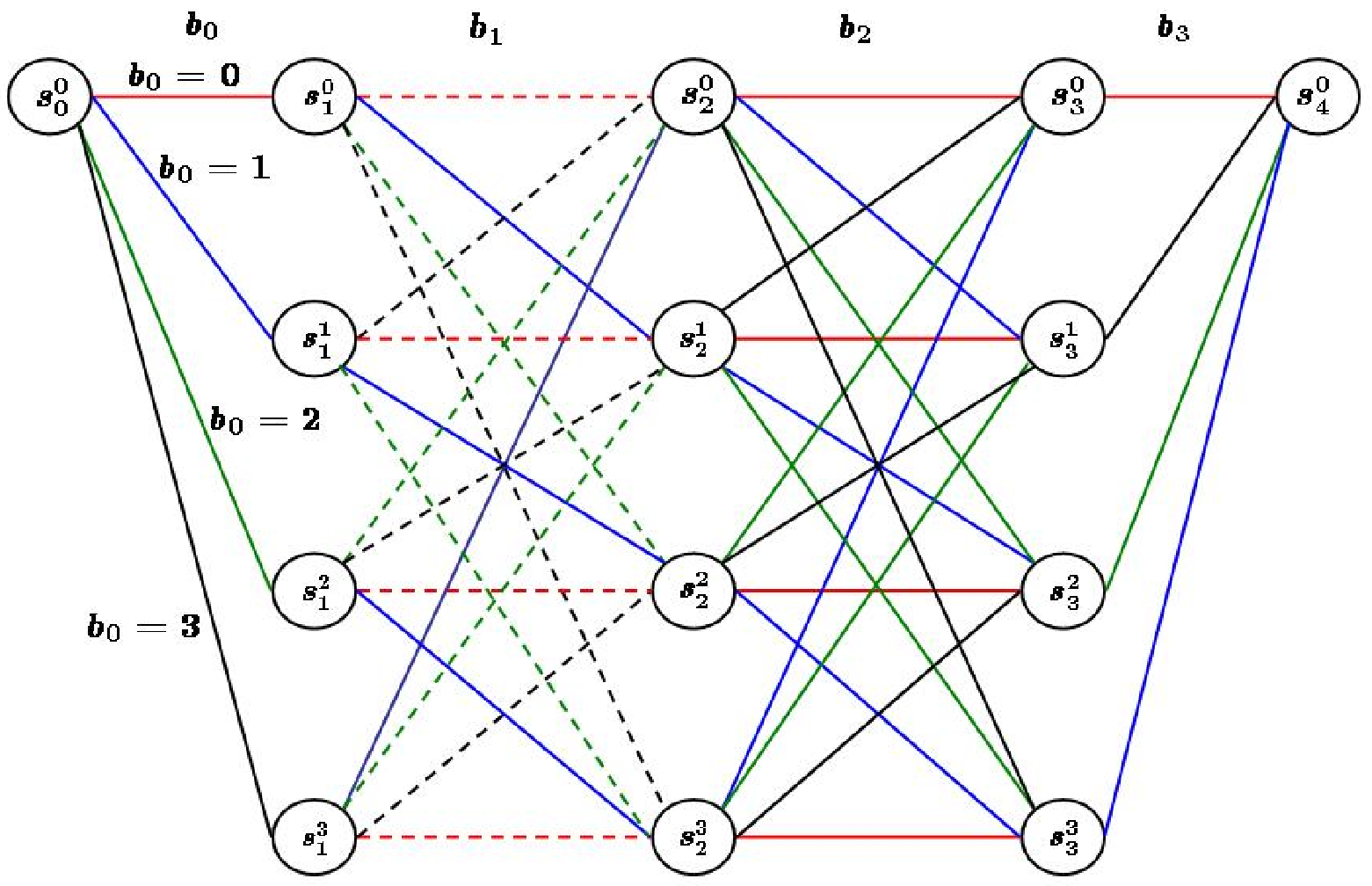}
\caption{States connected by dotted branches are used for the calculation of the $C_{j,\bar{1}}$.} 
\vspace{-15pt}
\label{fig:trellis_alpha_bar}
\end{minipage}
\end{figure*}
%
%
%
%
\begin{lemma} \label{lm:bcjr_compute}
$C_{j,{\alpha}}$ and $C_{j,\bar{\alpha}}$ can be efficiently computed on the trellis of the nonbinary code $\mathcal{C}_j$ as follows,
\begin{align}
& \sym \Big(C_{j,\bar{\alpha}}\Big) = \sum_{\underset{s'-s \ne {\alpha}} {(s, s') \in (\mathcal{S}_i, \mathcal{S}_{i+1})}} \mu_i({s}) \cdot \nu_{i+1}({s'}) \cdot g(s'-s) \label{eq:alpha_bar} \\
& \sym \Big(C_{j,\alpha}\Big) = \sum_{\underset{s'-s = \alpha} {(s, s') \in (\mathcal{S}_i, \mathcal{S}_{i+1})}} \mu_i({s}) \cdot \nu_{i+1}({s'}) \label{eq:alpha}
\end{align}
where state metrics $\mu_i$ and $\nu_{i+1}$ are calculated recursively from previous state metrics via
\begin{align*}
\mu_{i}(s) &= \sum_{{b_{i-1} \in \Re}} \mu_{i-1}({s-b_{i-1}}) \cdot g(b_{i-1}) \; , \\ 
\nu_i({s}) &= \sum_{\underset{}{b_{i} \in \Re}} \;\; \nu_{i+1}({s-b_{i}}) \cdot g(b_{i}) \; .
\end{align*}
\end{lemma}
\begin{proof}
First we prove that the state metrics can be computed recursively. The following may be observed from the definition of $\mu_i({s})$,
\begin{align*}
&\mu_i(s) = \sum_{\underset{\sigma(0,{i-1}) = s}{(b_0,\cdots,b_{i-1})}} \prod_{t = 0}^{i-1} g(b_t)\\
&= \sum_{\underset{\sigma(0, {i-2}) + b_{i-1} = s}{(b_0,\cdots,b_{i-1})}} \left(\prod_{t = 0}^{i-2} g(b_t)\right) \cdot g(b_{i-1})\\
&= \sum_{\underset{}{b_{i-1} \in \Re}} \left( \sum_{\underset{\sigma(0, i-2) = s-b_{i-1}}{(b_0,\cdots,b_{i-2})}} \prod_{t = 0}^{i-2} g(b_t)\right) \cdot g(b_{i-1})\\
&= \sum_{\underset{}{b_{i-1} \in \Re}} \mu_{i-1}({s-b_{i-1}}) \cdot g(b_{i-1})
\end{align*}
Hence $\mu_i(s)$ can be calculated recursively from the previous state metrics. Similarly, we can prove that the $\nu_{i}({s})$ can be calculated from previous state metrics.

Now we prove the other part of the lemma. For ease of exposition we assume $\mathcal{I}_j =\{0,\cdots,d_j-1\}$ in the following. 
\begin{align*}
\sym \Big(C_{j,\bar{\alpha}}\Big) &= \sum_{\underset{b_i \ne \alpha}{\bs{b} \in \mathcal{C}_j}} \sym \Big(\langle \hat{\bs{v}}_j, \Xi({\bs{b}}) \rangle\Big) \nonumber \\
&= \sum_{\underset{b_i \ne \alpha}{\bs{b} \in \mathcal{C}_j}} \sym \left( \sum_{{t = 0}}^{d_j-1} \Big( \langle \hat{\bs{v}}_{j,t}\;,\; {\xi(b_t)}\rangle\Big) \right) \nonumber\\ 
&= \sum_{\underset{b_i \ne \alpha}{\bs{b} \in \mathcal{C}_j}} \left( \prod_{{t = 0}}^{d_j-1} \sym \Big( \langle \hat{\bs{v}}_{j,t}\;,\; {\xi(b_t)}\rangle \Big) \right) \nonumber
\end{align*}
\begin{align}
&\Rightarrow \sym \Big(C_{j,\bar{\alpha}}\Big) = \sum_{\underset{b_i \ne \alpha}{\bs{b} \in \mathcal{C}_j}} \left( \prod_{{t = 0}}^{d_j-1} g(b_t) \Big) \right) \label{eq:lc_lp}
\end{align}
The right-hand side of \refeqwo{eq:alpha_bar} is,
\begin{align}
&\sum_{\underset{s'-s \ne \alpha} {(s, s') \in (\mathcal{S}_i, \mathcal{S}_{i+1})}} \mu_i({s}) \cdot \nu_{i+1}({s'}) \cdot g(s'-s) \nonumber \\
&= \sum_{\underset{s'-s \ne \alpha} {(s, s') \in (\mathcal{S}_i, \mathcal{S}_{i+1})}} \left(\sum_{\underset{\sigma(0, {i-1}) = s}{(b_0,\cdots,b_{i-1})}} \prod_{t = 0}^{i-1} g(b_t)\right) \nonumber \\
&\quad\quad\quad\quad \left(\sum_{\underset{\sigma(i+1,d_j-1) = s'}{(b_{i+1},\cdots,b_{d_j-1})}} \; \prod_{t=i+1}^{d_j-1} g(b_t)\right) \cdot g(b_i) \nonumber \\
&= \sum_{\underset{s'-s \ne \alpha} {(s, s') \in (\mathcal{S}_i, \mathcal{S}_{i+1})}} \sum_{\underset{\sigma(0, {i-1}) = s, \sigma(i+1,d_j-1) = s'}{(b_0,\cdots,b_{i-1}, b_{i+1},\cdots,b_{d_j-1})}}  \nonumber\\
&\quad\quad\quad\quad\quad \left( \prod_{t = 0}^{i-1} g(b_t) \cdot \prod_{t=i+1}^{d_j-1} g(b_t) \right) \cdot g(b_i) \nonumber
\end{align}
\begin{align}
\Rightarrow \sum_{\underset{s'-s \ne \alpha} {(s, s') \in (\mathcal{S}_i, \mathcal{S}_{i+1})}} &\mu_i({s}) \cdot \nu_{i+1}({s'}) \cdot g(s'-s) \nonumber \\
&= \sum_{\underset{b_i \ne \alpha}{\bs{b} \in \mathcal{C}_j}}  \left( \prod_{{t = 0}}^{d_j-1} g(b_t)\right) \label{eq:bcjr}
\end{align}
Using \refeqwo{eq:bcjr} in \refeqwo{eq:lc_lp} we get \refeqwo{eq:alpha_bar}. Equation \refeqwo{eq:alpha} can be proved in a similar manner.
\end{proof}

The overall algorithm works in two phases: in the first phase, the forward and backward state metrics are calculated and stored; in the second phase the marginals $C_{j,{\alpha}}$  and $C_{j,\bar{\alpha}}$ are computed with \reflem{lm:bcjr_compute} where the state metrics computed in first phase are utilized. It may be observed that the aforementioned algorithm is essentially the same as the BCJR algorithm except for the second phase where marginals are calculated.

The calculations of the \reflem{lm:bcjr_compute} can be visualized with the help of the trellis diagram. Figures \reffigwo{fig:trellis_alpha} and \reffigwo{fig:trellis_alpha_bar} shows the trellis for the nonbinary SPC code of length $4$ which is defined over $\mathbb{Z}_4$. $b_0$ to $b_3$ represent the symbols, and states are represented by $s_{t}^{i}$, where $t$ indicates the symbol after which the state occurs and $i$ represents the sum of the symbols from $b_0$ to $b_{t-1}$. The dotted branches in \reffig{fig:trellis_alpha} represents the transitions related to the symbol $b_1 = 1$. The state pairs which are connected by these branches are used for the calculation of the $C_{j,1}$. Similarly, the dotted branches in \reffig{fig:trellis_alpha_bar} represent transitions related to the symbol $b_1 \ne 1$. Here the metrics of the corresponding state pairs are used for the calculation for the $C_{j,\bar{1}}$.
\subsection{Alternative State Metric for Faster Calculation of $C_{j,\bar{\alpha}}$}
The forward state metric $\mu$ as defined in \refeqwo{eq:mu} needs to be computed for the calculation of $C_{j,\alpha}$ and can be reused for the calculation of $C_{j,\bar{\alpha}}$. In \refeqwo{eq:alpha_bar} the algorithm needs to go through all branches $(s, s') \in (\mathcal{S}_i, \mathcal{S}_{i+1}), s' - s \ne \alpha$ for the calculation of $C_{j,\bar{\alpha}}$. If the proposed algorithm is implemented in hardware or on multicore architectures, then the computation time for $C_{j,\bar{\alpha}}$ can be reduced by parallelizing its calculation. One possibility to parallelize calculation of $C_{j,\bar{\alpha}}$ is to define a new forward state metric $\bar{\mu}$, which can be computed in parallel with $\mu$ in the first phase and reduces the calculations required during the second phase of the algorithm. For this we define an alternative forward state metric as follows,
\begin{align}
\bar{\mu}_i({s,{\alpha}}) = \sum_{\underset{\sigma(0, {i-1}) = s, b_{i-1} \ne \alpha}{(b_0,\cdots,b_{i-1})}} \prod_{t = 0}^{i-1} g(b_t), \;\; s \in \mathcal{S}_i, i \in \mathcal{I}_j, \alpha \in \Re^{-} \label{eq:new_mu}
\end{align}
\vspace{-15pt}
\begin{align*}
\text{ with } \; \bar{\mu}_{0}({s,\alpha}) = 0, \; \forall s \in \mathcal{S}_0, \; \forall \alpha \in \Re^{-}. \nonumber
\end{align*}
It should be noted that due to the condition $b_{i-1} \ne \alpha$, $\bar{\mu}_{i}({s,{\alpha}})$ cannot be calculated recursively from $\bar{\mu}_{i-1}$; instead it is calculated together with $\mu_{i}({s})$ from $\mu_{i-1}$ as follows,
\begin{align*}
&\bar{\mu}_{i}({s,{\alpha}}) = \sum_{\underset{}{b_i \in \Re\setminus\{\alpha\}}} \mu_{i-1}({s-b_i}) \cdot g(b_i)
\end{align*}
With the help of the alternative forward state metric given in \refeq{eq:new_mu}, the expression \refeqwo{eq:alpha_bar} of \reflem{lm:bcjr_compute} can be rewritten as
\begin{align}
\sym\Big(C_{j,\bar{\alpha}}\Big) = \sum_{s' \in \mathcal{S}_{i+1}} \bar{\mu}_{i+1}({s',{\alpha}}) \cdot \nu_{i+1}(s') \label{eq:new_alpha_bar}
\end{align}

The forward state metric $\bar{\mu}_{i}({s,\alpha})$ requires the calculation and storage of an additional $q-1$ values for each state $s \in \mathcal{S}_i$ during the first phase. Hence the storage requirement for the calculation of $C_{j,\bar{\alpha}}$ with \refeqwo{eq:new_alpha_bar} increases by a factor of $q$.
However, all additional state metric values can be calculated in parallel with $\mu$ which does not effect the run time of the first phase of the algorithm. Also, the second phase of the algorithm needs to go through only $q$ states instead of $q(q-1)$ branches, hence the overall run time for computing $C_{j,\bar{\alpha}}$ is reduced with the state metric $\bar{\mu}$.
%
%
%
%
%
\subsection{Calculation of Marginals with $\kappa \to \infty$}
In \reflem{lm:bcjr_compute}, $\kappa$ is assumed to be finite. However, for many practical applications we are interested in $\kappa \to \infty$. According to Lemma 6.3 of \cite{PuFl_10}, for $\kappa \to \infty$ we again need to calculate $(C_{j,\alpha} - C_{j,\bar{\alpha}})$ to update the corresponding variables.
However, the marginals $C_{j,\alpha}$ and $C_{j,\bar{\alpha}}$ are here obtained as the limit of \refeq{eq:c_alpha} and \refeqwo{eq:c_alpha_bar} respectively as $\kappa \to \infty$, i.e.,
\begin{equation}
C_{j,\alpha} \triangleq - \min_{\underset{b_i = \alpha}{\bs{b} \in \mathcal{C}_j}} \langle -\tilde{\bs{v}}_j, \Xi(\tilde{\bs{b}}) \rangle,\; C_{j,\bar{\alpha}} \triangleq - \min_{\underset{b_i \ne \alpha}{\bs{b} \in \mathcal{C}_j}} \left\langle -\hat{\bs{v}}_j, \Xi(\bs{b}) \right\rangle \label{eq:new_c}
\end{equation}
Thus $C_{j,{\alpha}}$ and $C_{j,\bar{\alpha}}$ can be obtained by replacing all ``product" operations with ``sum" operations and similarly by replacing all ``sum" operations with ``min" operations in \refeqwo{eq:c_alpha} and \refeqwo{eq:c_alpha_bar} (marginals with finite $\kappa$). In \refeqwo{eq:c_alpha} and \refeqwo{eq:c_alpha_bar} the marginalization is performed in the sum-product semiring. However for $\kappa \to \infty$ the marginalization is performed in the min-sum semiring and hence the marginals of \refeqwo{eq:new_c} can be computed with a trellis based variant of the min-sum algorithm. If we redefine the branch metric as $g(b_t) = \langle\hat{\bs{v}}_{j,i}\;,\;\xi(b_t)\rangle$ and replace all ``product" operations with ``sum" operations and similarly replace all ``sum" operations with ``min" operations in \refeq{eq:mu}, \refeqwo{eq:nu}, \refeqwo{eq:alpha}, \refeqwo{eq:new_mu} and \refeqwo{eq:new_alpha_bar} then the resulting equations can be used on the trellis of the nonbinary SPC code $\mathcal{C}_j$ to compute the marginals of \refeqwo{eq:new_c}. This trellis based variant of the min-sum algorithm is related to the Viterbi algorithm.
\begin{figure}
\centering
\includegraphics[width=1.0\columnwidth, keepaspectratio]{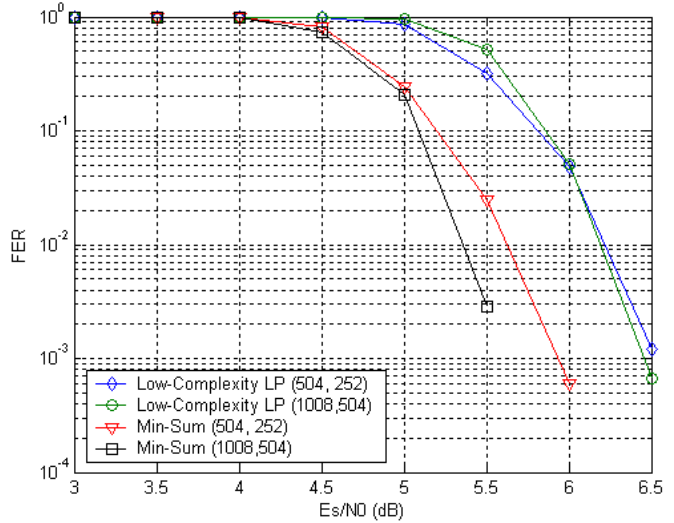}
\caption{Frame Error Rate for $(504, 252)$ and $(1008, 504)$ quaternary LDPC code under QPSK modulation. The performance of Low complexity LP decoding is compared with that of the min-sum algorithm.} 
\vspace{-15pt}
\label{fig:FER_all}
\end{figure}
%
%
%
%
%
%
\section{Results}
This section presents simulation results for low complexity LP decoding which uses the trellis based check node calculations described above. We consider $\kappa \to \infty$ for all simulations. We use the binary $(504, 252)$ and $(1008, 504)$ MacKay LDPC codes, but with parity-check matrix entries taken from $\mathbb{Z}_4$ instead of $GF(2)$. These LDPC codes are $(3,6)$-regular codes; hence there are $6$ nonzero entries in each row of their parity-check matrix. We set the second and third nonzero entry in each row to $3$, and all other nonzero entries are set to 1. Furthermore, we assume transmission over the AWGN channel where nonbinary symbols are directly mapped to quaternary phase-shift keying (QPSK) signals. We simulate up to $100$ frame errors per simulation point.

The error-correcting performance of the $(504, 252)$ and $(1008, 504)$ LDPC code is shown in \reffig{fig:FER_all} where the frame error rate (FER) of the LCLP decoding algorithm is compared with that of the min-sum (MS) algorithm. The MS algorithm also uses the trellis of the nonbinary SPC code for check node processing. The maximum number of iterations is set to $64$ for both decoding algorithms. For the $(504, 252)$ code, the FER of low complexity LP decoding is within $0.5$ dB from that of MS algorithm and for $(1008, 504)$ code, it is within $0.7$ dB. These results are comparable to that of the binary LCLP decoding algorithm of \cite{VoKo_06}. Finally, it is important to note that these LDPC codes are significantly longer then the quaternary $\text{(80, 48)}$ LDPC code tested in \cite{PuFl_10}.
%
%
%
\section{Conclusion}
In this paper, we proposed a modified BCJR algorithm for efficient check node processing in the nonbinary LCLP decoding algorithm. The proposed algorithm has complexity linear in the check node degree. We also proposed an alternative state metric which can be used to reduce the run time of the proposed algorithm.
\section{ACKNOWLEDGMENTS}
The authors would like to thank P. O. Vontobel for many helpful suggestions and comments. This work was supported in part by the Claude Shannon Institute, UCD, Ireland.


\begin{thebibliography}{99}
%
\bibitem{DaMa_98}
M. C. Davey and D. J. C. MacKay, ``Low density parity check codes over $GF(q)$," \emph{IEEE Communication Letters}, vol. 2, no. 6, pp. 165--167, June 1998.
%
\bibitem{FeWa_05}
J. Feldman, M.~J. Wainwright and D.~R. Karger, ``Using linear programming to decode binary linear codes," \emph{IEEE Transactions on Information Theory}, vol. 51, no. 3, pp. 954--972, March 2005.
%
\bibitem{VoKo_06}
P. O. Vontobel and R. Koetter, ``Towards low-complexity linear-programming decoding," in \emph{Proc. of 4th International Conference on Turbo Codes and Related Topics}, Munich, Germany, April 3--7, 2006.
%
 
\bibitem{FlSk_09}
M. F. Flanagan, V. Skachek, E. Byrne, and M. Greferath, ``Linear-Programming Decoding of Nonbinary Linear Codes," \emph{IEEE Transactions on Information Theory}, vol. 55, no. 9, pp. 4134--4154, September 2009.
%
\bibitem{PuFl_10}
M. Punekar and M. F. Flanagan, ``Low Complexity LP Decoding of Nonbinary Linear Codes," The \emph{Forty-Eighth Annual Allerton Conference on Communication, Control, and Computing}, September 29 -- October 1, 2010.
%
\end{thebibliography}
\end{document}